\documentclass[9pt,twocolumn,twoside]{osajnl}
\usepackage{graphicx}

\journal{ol} 

\setboolean{shortarticle}{true}

\title{Study of adhesivity of surfaces using rotational optical tweezers}

\author[1]{Rahul Vaippully}
\author[1]{Dhanush Bhatt}
\author[1]{Anand Dev Ranjan}
\author[1]{Basudev Roy}

\affil[1]{Physics Department, Indian Institute of Technology Madras, Chennai, 600036}

\affil[*]{Corresponding author: basudev@iitm.ac.in}

\begin{abstract}
Optical tweezers are powerful tools for high resolution study of surface properties. Such experiments are traditionally performed by studying the active or the brownian fluctuation of trapped particles in the X, Y, Z direction. Here we find that employing the fourth dimension, rotation, allows for sensitive and fast probing of the surface. Optical tweezers are capable of rotating trapped birefringent microparticles when applied with circularly polarized light, thus called the Rotational Optical Tweezers. When the trapped birefringent microparticle is far enough away from the surface, the rotation rate is dependent only on the laser power. However, we find that if one traps close to a surface, the rotation rate goes to zero even at finite tweezers laser powers for some specific type of substrates. We suspect this to be due to interaction between the substrate and the birefringent particle, keeping in mind that the hydrodynamic drag for this mode of rotation cannot increase beyond 1.2 times the drag away from the surface. We use this to probe some surfaces and find that there is no binding for hydrophobic ones but hydrophilic ones particularly tend to show a power threshold beyond which the birefringent particle starts rotating. We calculate that the threshold energy of the tweezers is consistent with the Van der Waals potential energy, when the mode of interaction with the surface is purely physical. We also find that for chitosan, the mode of interaction is possibly different from Van der Waals. We place the particle on the threshold and observe ”stick-slip” kind of rotational behaviour.
\end{abstract}

\begin{document}

\maketitle

The study of adhesivity at the microscopic scales has generally been performed using Atomic Force Microscopes (AFM) \cite{florin}. This applies a nanonewton force and can also detect interactions to a minimum of about 100 pN nm due to limitations in the cantilever size \cite{tilman}. Further, hard probing of live biological cells using AFM probes can rupture the membrane thereby killing them. Here we present an alternative technique using birefringent probes trapped in optical tweezers \cite{halina} which can probe softly \cite{rohrbach} and sense interactions at higher resolution. Since we do not apply a large normal force on the surface, there is no possiblity of rupturing membranes of live cells either. 

Rotational motion of a birefringent microsphere using an circularly polarized optical tweezers apparatus has been described with the following eq. \ref{eq1} (\cite{halina})
\begin{equation}
\gamma \frac{d\theta}{dt} = \tau = \eta I
\label{eq1}
\end{equation}

where, $\gamma$ is the drag coefficient, $\tau$ the rotational torque due to circularly polarized light, $\eta$ the efficiency of torque transfer and I is the intensity of the tweezers light. Here, we have assumed that the gaussian varying random noise due to Brownian motion is negligible compared to the rotational torque. 

We can generally conclude that the rotational rate would go to zero only when the laser intensity would be brought to zero. An interesting case appears when we bring this rotating microsphere close to a surface.  The rotational faxen correction to the drag is bounded to a value ($\zeta$(3) which is about 1.20, where $\zeta$ indicates the Riemann zeta function) very close to the surface while being compared to the value away from surface \cite{liu,padgett}, thereby indicating that the rotation rate can never become zero at a finite value of torque.
 We perform our measurements close to some mildly hydrophilic surfaces using birefringent liquid crystalline RM257 microspheres \cite{avin} and find that in surface proximities, the particle stops rotating even at a finite value of torque. We suspect that this is an effect of adhesion to the surface, which must be overcome for rotation. 
A similar study was carried out for translational mode of motion of a particle moving parallel to a surface when the rheology of sticking transition was studied \cite{prerna}. However, this mode of translation faces much larger Faxen corrections than the yaw rotational one \cite{pitch}, not to mention the time scales of interaction inviting slower effects.

\begin{figure}[htbp]
\centering
\includegraphics[width=\linewidth]{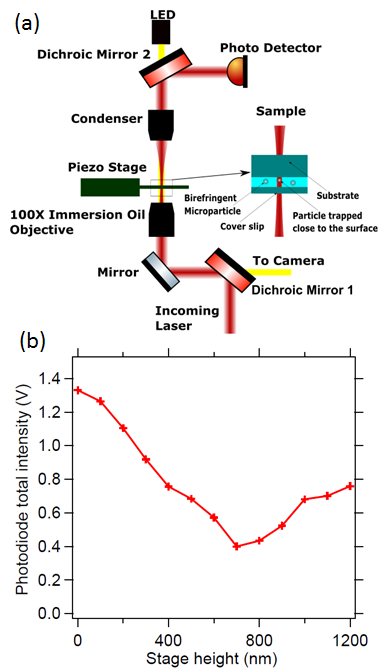}
\caption{(a) Schematic diagram showing the optical tweezer set up. The trapping is made close to the top surface of the sample chamber. (b) Change in the total intensity of the scattered light as the particle is brought closer to the surface using a translation stage in steps of 100 nm. The position of the minimum of the scattered light signal (the point of inflection) indicates an axial distance of of 100 nm from the surface. Here 0 nm indicates a position futher away from the surface than 1200 nm. }
\label{schematic}
\end{figure}

In order to perform the measurement, we optically trap and rotate a birefringent microparticle (liquid crystalline colloid of RM257 material (Merck), chemical name 2-Methyl-1,4-phenylene bis(4-(3-(allyloxy)propoxy)benzoate) \cite{avin,basudev}, of diameter 1 $\pm$ 0.2 $\mu$m, typical birefringence of 0.01) close to the top surface in an inverted microscopy configuration using the optical Tweezers Kit (Thorlabs USA) as shown in Fig. \ref{schematic}. Colloids of RM257 are stable in water. The illumination objective is 1.25 NA, E Plan 100x objective from Nikon at the bottom with the illumination aperture being overfilled. The collection objective is E Plan 10x, 0.25 NA also from Nikon. The laser used for optical trapping is 976 nm Butterfly laser (Thorlabs, USA). The particles are trapped close to the top surface of the sample chamber and all the measurements performed in this configuration. We collect all the forward scattered light incident on a photodiode and move the particle close to the surface. The scatter intensity keeps on reducing till it reaches a minimum, whereafter it keeps increasing. The position of the surface when the inflection point of forward scattered light is attained is known to be about 100 nm from the surface, given in Fig. 5(a) of \cite{erik}, and also shown in Fig. \ref{schematic}. The slopes of the lines before and after the surface is reached is noticably different, from where the surface position can be inferred. As the surface is reached, where the inflection point is attained, the particle starts to get pushed away from the equilibrium of the trap upon the effect of the surface, thereby showing a different slope. 

\begin{figure}[htbp]
\centering
\includegraphics[width=\linewidth]{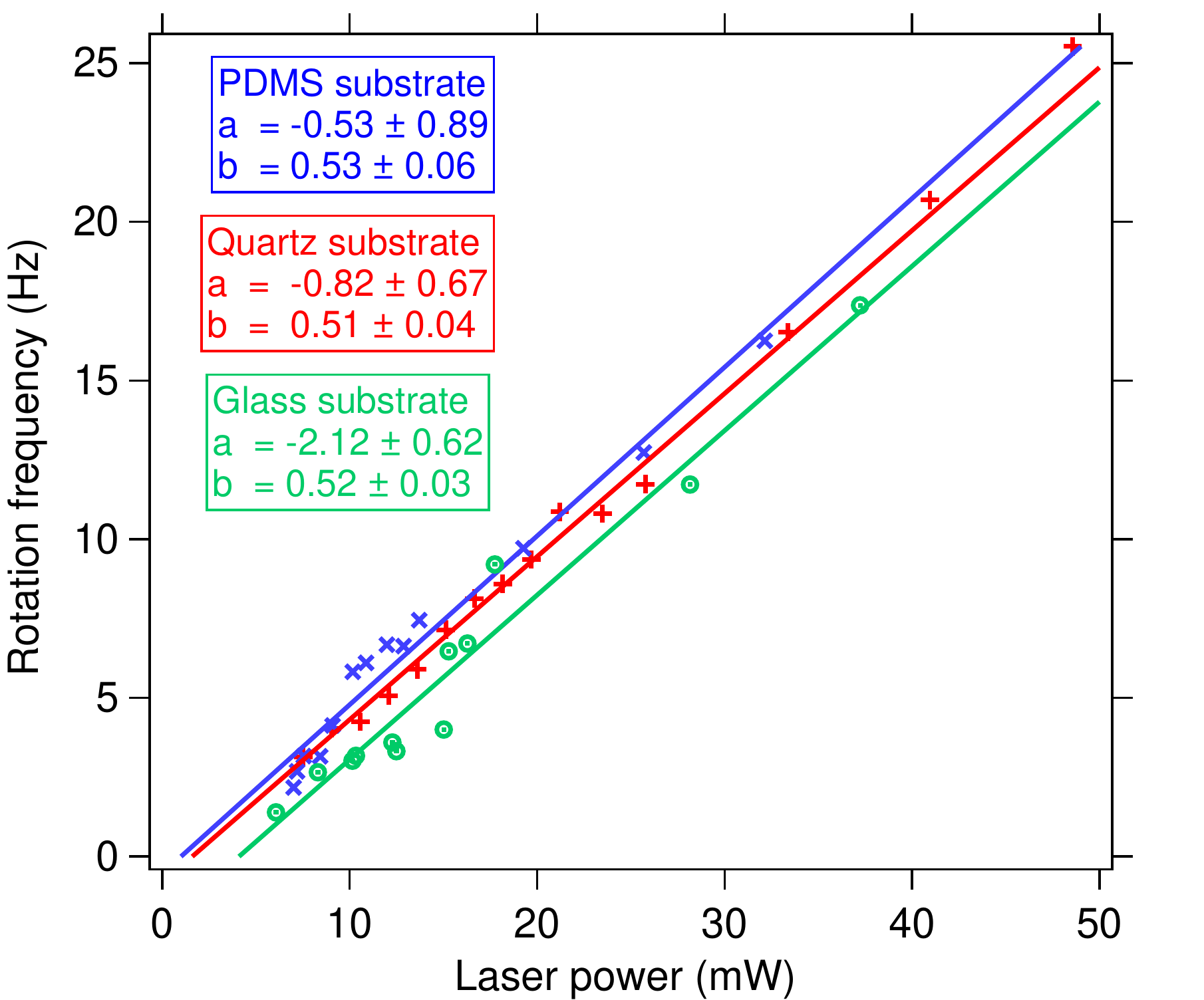}
\caption{Rotation rate as a function of laser power for PDMS, Quartz \& Glass Surfaces. Quartz and glass seem to have a threshold while PDMS shows no threshold. The data have been fitted to an equation of the form y = b x + a. }
\label{powervsfreq}
\end{figure}

When we trap the particle close to the glass surface and gradually reduce the laser power, we find a threshold beyond which the particle stops rotating.
It has been shown in Fig. \ref{powervsfreq}. There is however no such threshold if a PDMS substrate is used. We investigate this further by considering that PDMS is hydrophobic in nature while the glass surface that we have used is hydrophilic. We put droplets of water onto the glass slide and estimate the contact angle which we find to be about 45 degrees. We also consider a quartz slide that shows a contact angle of 75 degrees, indicating that it is less hydrophilic. The respective rotation rate as a function of laser power curve has been shown on Fig. \ref{powervsfreq}. In all these measurements for rotation rate as a function of laser power, we reduce the power till the rotation rate is about 1 Hz under which the trapping becomes so weak that the vertical excursions from equilbrium are comparable to the rotation events what we are trying to detect. Thus reducing laser power under such values do not give trustworthy results. Further, considering that the suspensions of RM257 form stable colloids in water and that these tend to stick to the hydrophilic subtrates tends to indicate that this particle itself is  hydrophilic in nature.  

The hydrophobic surface hardly has a threshold, as the straight line passes through the origin within the error bar of the fit. However, there are noticable thresholds for the quartz and the regular glass slide, with the quartz having a lower threshold than the glass slide. This can be explained by the following equation

\begin{equation}
\gamma \frac{d\theta}{dt} = \eta (I - I_0)
\label{eq2}
\end{equation}

where, $I_0$ is the threshold that needs to be overcome to initiate rotation. This can be understood as bonds being formed between the probe and the surface, which requires $\eta$ $I_0$ amount of torque to be overcome. The slope of line is $\frac{\eta}{\gamma}$ which for similar types of particles would be same since $\eta$ only depends upon the particle size and birefringence of the particle, while the $\gamma$ depends upon the viscosity of the medium and the size of the particle. We do indeed find the slopes to be same for the three different surfaces since the sizes of all of them are about 1 micron. 

\begin{figure}
\centering
\includegraphics[width=\linewidth]{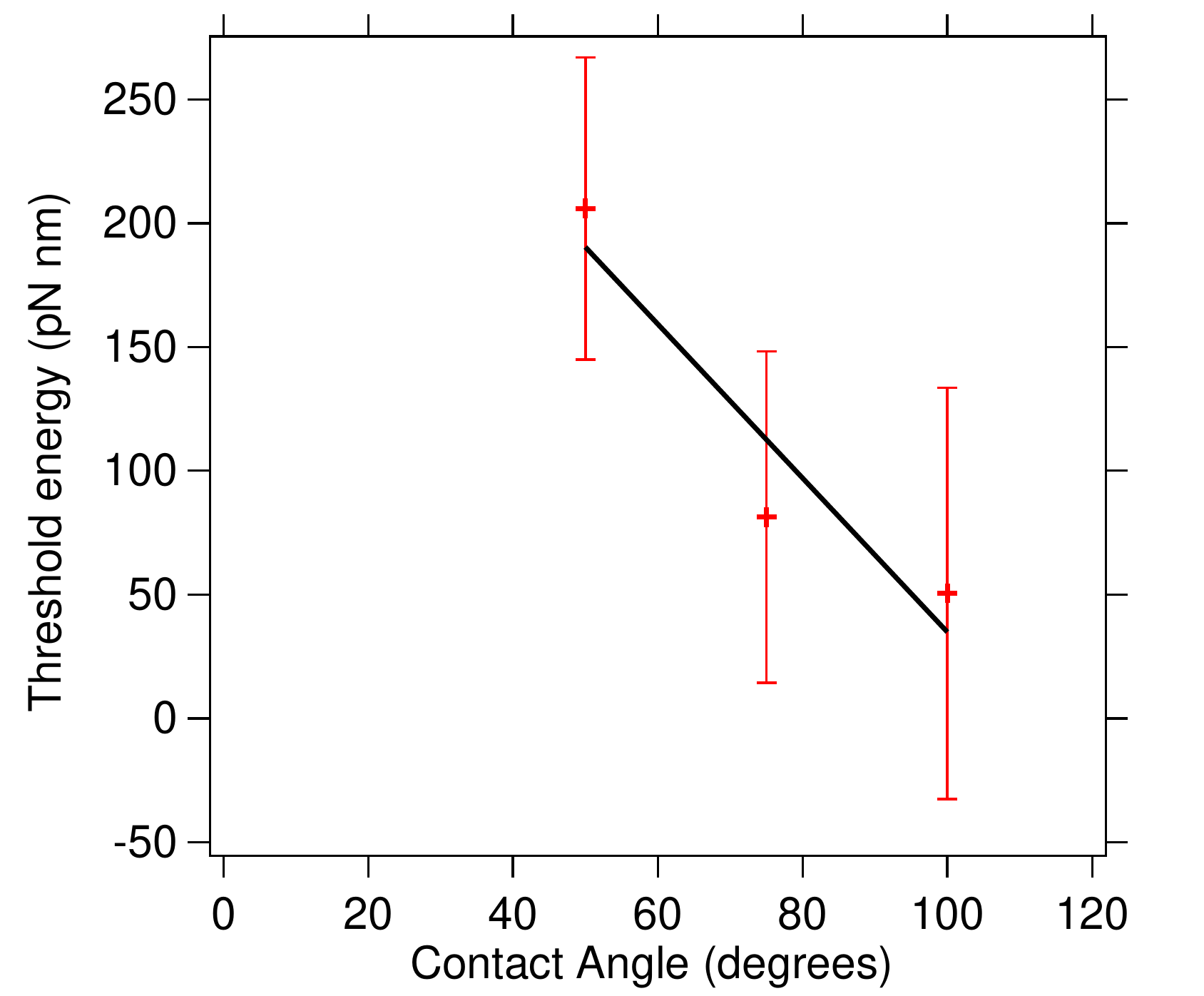}
\caption{Threshold laser energy to initiate rotation as function of a water drop contact angle on the substrate. The energy required to initiate rotation seems to indicate van der waals potentials as the binding mechanism to the surface. }
\label{vdw}
\end{figure}

We compute the corresponding threshold energy that these particles must overcome to initiate rotation. The threshold power for rotation on glass is about 4.25 mW and the corresponding threshold a = -2.12 Hz. The drag coefficient $\gamma$ = 8 $\pi$ $\eta_1$ r$^3$, where $\eta_1$ is the viscosty of water and r is the radius of the particle. Then the coefficient $\eta$/$\gamma$ = 0.52 $\times$(2$\pi$)) Hz/mW, such that the value of $\eta$ = 8.05 $\times$ 10$^{-21}$ pN nm Hz/mW. Then, the threshold torque is $\eta \times I_0$ which is 34.2 pN nm and the subsequent energy is to rotate it by 360 degrees which is 34.2 pN nm $\times$ 2$\pi$ = 215 pN nm. We find them to be of the order of 50-220 pN nm, as shown in Fig. \ref{vdw}. These have been plotted as a function of contact angle of a water droplet on the surface and exhibit a straight line fit within error bars. The threshold energy corresponds very well to the Van der Waals interaction \cite{hamaker} with a Hamaker constant of about 2 $\times$ $10^{-27}$ J m separated from the particle to the surface by less than 100 nm. Thus, the rotation threshold is a good measure of the hydrophilicity of the substrate given that the mechanism of interaction is Van der Waals interaction.

\begin{figure}
\centering
\includegraphics[width=\linewidth]{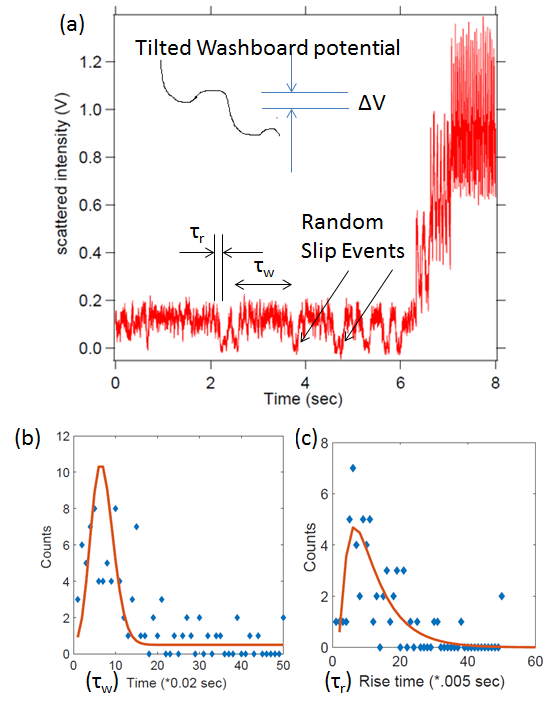}
\caption{(a) Time series for rotation at four different laser powers on a glass substrate. At the minimum laser power used here, the rotational events become random and can be described as slip events. (b) Time distribution of slip events fitted to a Poisson distribution. (c) Distribution of rise times for the slip events fitted to a curve indicated in \cite{kall}, expected for transitions in a biased double well potential. This gives the height of the bistable potential well. }
\label{fig5}
\end{figure}

We show a typical curve for rotational motion of the microsphere close to the surface in Fig. \ref{fig5} (a). At high values of laser power, the particle continues to rotate periodically. However, as the power is reduced slowly, the rotation rate also reduces till a point where the rotational events start to become random. When we study waiting time distribution for the delay between two slip events, we find a distribution of the form Fig. \ref{fig5}(b) which can be fitted well to a Poisson distribution. We also study the time it takes for the rotation of the particle by 180 deg during the slip events which follows a distribution given in \cite{kall}, shown in Fig. \ref{fig5}(c) and expected for a biased double well potential. A tilted washboard potential can be also referred to a biased double well potential due to rotational symmetry. We estimate that the bistable potential has a barrier ($\Delta$V) of the height ~ 0.8 $k_B$ T, quite consistent with other types of such jumps in a tilted washboard potential \cite{basudev,kall,pedaci,pedaci2}. 

\begin{figure}
\centering
\includegraphics[width=0.3\textwidth]{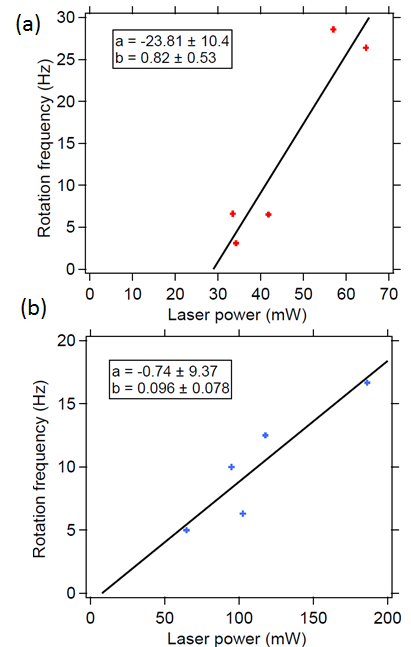}
\caption{Rotation rate as a function of laser power for (a) Chitosan substrate and (b) CHO cell surface. The chitosan sample seems to exhibit a proper threshold while the CHO cell seems have no threshold with a much lower slope than the harder surfaces. The data have been fitted to a form y = b x + a. }
\label{fig4}
\end{figure}

We go on to study a different type of substrate where the mechanism of interaction is not expected to be of Van der Waals type, namely that of chitosan. We find that even though the contact angle for the chitosan substrate is about 75 degrees, the threshold is much larger than that of glass, at 1400 pN nm. This can be explained by the fact that chitosan is a well known bio-adhesive with stickiness extending to a pH of 7 \cite{chitosan}. We believe to have recorded this in the threshold of rotation. We show this Fig. \ref{fig4}(a).  

We also tried to perform the rotation close to a Chinese Hamster Ovary (CHO) cell which was prepared to be adhered to a glass substrate and find a very different behavior, as shown in Fig. \ref{fig4}(b). The slope of the curve seems to be much smaller (0.09 Hz/mW compared to about 0.51 Hz/mW in Fig. \ref{powervsfreq}) while the threshold seems to be absent. We can explain this by considering that the threshold $\eta I_0$ indicated in eq. \ref{eq2} is dependent on frequency as $\gamma_0 \omega$. 

\begin{equation}
\gamma \omega = \eta I - \gamma_0 \omega
\label{eq3}
\end{equation}

Then we get the following equation where the slope can be much lower. 

\begin{equation}
(\gamma + \gamma_0) \omega = \eta I 
\label{eq4}
\end{equation}
Although the viscous drag coefficient depends upon the particle size, the RM257 sample diameter is monodisperse to within 1000 $\pm$ 200 nm \cite{avin}, also confirmed with video imaging. This does not explain the low slope. This indicates that the CHO cell surface appears like a viscous medium for the rotating particle with a viscosity which is 5 times that of a particle in proximity to glass. Since the trapping light enters the sample chamber from the bottom while the CHO has been attached to the substrate at the top, the cell itself has no effect on the trapping light. This kind of effect has also be reported in a similar work using translational motion close to the surface upon the influence of specific binding to the surface \cite{pralle}. We suspect that our increase in drag is due to partial nonspecific binding to the surface of the cell.  

Thus, to conclude, we have demonstrated a system using birefringent microspheres which can sense the surface adhesivity. If surfaces without chemical adhesivity are used, the Van der Waals force ensures that only hydrophilic ones show rotational threshold. This can subsequently be used to determine the hydrophilicity of the surface using a hydrophilic probe. Further, even if other mechanisms of binding are present, the rotational threshold indicates the binding energy. We also show that at the point of threshold, particularly on the glass surfaces which are mildly hydrophilic, the rotational motion follows a stick-slip behaviour where the slip events happen randomly. This kind of technique can eventually be used to study adhesivity at the nanometric scales and is possibly a more sensitive probe than Atomic Force Microscopes.

We thank the Indian Institute of Technology Madras for the seed grant. We also thank Amal Kanti Bera for providing us the CHO cells required for the experiment, Dillip Satapathy for the PDMS and Chitosan sample and Erik Schaffer for the RM257 liquid crystal powder. 

\subsection{Materials and Methods}
The chitosan (Sigma-Aldrich 50000) substrate was prepared by dissolving in  5\% solution of formic acid in double distilled water. We took 3\% (W/V) concentration of chitosan in the formic acid solution for making the film. This solution is stirred at 75$^{\circ}$C for one hour. The prepared solution is filtered and spin coated on glass substrate. \\  
The PDMS (Dow Corning's Sylgard 184 elastromer kit) substrate is prepared by curing the Silicone elastomer in 10\% (W/W) with the curing agent. The viscous solution is spin coated on glass slides and baked at 150$^{\circ}$C.

\bibliography{sample}

\bibliographyfullrefs{sample}


\ifthenelse{\equal{\journalref}{aop}}{%
\section*{Author Biographies}
\begingroup
\setlength\intextsep{0pt}
\begin{minipage}[t][6.3cm][t]{1.0\textwidth} 
  \begin{wrapfigure}{L}{0.25\textwidth}
    \includegraphics[width=0.25\textwidth]{john_smith.eps}
  \end{wrapfigure}
  \noindent
  {\bfseries John Smith} received his BSc (Mathematics) in 2000 from The University of Maryland. His research interests include lasers and optics.
\end{minipage}
\begin{minipage}{1.0\textwidth}
  \begin{wrapfigure}{L}{0.25\textwidth}
    \includegraphics[width=0.25\textwidth]{alice_smith.eps}
  \end{wrapfigure}
  \noindent
  {\bfseries Alice Smith} also received her BSc (Mathematics) in 2000 from The University of Maryland. Her research interests also include lasers and optics.
\end{minipage}
\endgroup
}{}

\end{document}